\documentclass[graybox]{svmult}

\usepackage{natbib}
\usepackage{mathptmx}       % selects Times Roman as basic font
\usepackage{helvet}         % selects Helvetica as sans-serif font
\usepackage{courier}        % selects Courier as typewriter font
\usepackage{graphicx}        % standard LaTeX graphics tool
\newcommand{\aap}{A\&A}

\newcommand{\apj}{ApJ}
\newcommand{\apjl}{ApJL}
\newcommand{\aj}{AJ}
\newcommand{\araa}{ARA\&A}
\newcommand{\mnras}{MNRAS}

\newcommand{\nat}{Nature}

\newcommand{\kpc}{{\rm kpc}}

\newcommand{\Myr}{{\rm Myr}}

\newcommand{\lcdm}{$\Lambda$CDM}
\newcommand{\kms}{\;{\rm km}\,{\rm s}^{-1}}
\newcommand{\msun}{M_{\odot}}

\begin{document}

%\title*{The Impact of Gas Accretion on Galactic Chemical Evolution: Theory and Observations}
\title*{Gas Accretion and Galactic Chemical Evolution: Theory and Observations} %abbrieviated; check with Kristian
\author{Kristian Finlator}
\institute{Kristian Finlator \at New Mexico State University, Las Cruces, NM, USA \at \email{finlator@nmsu.edu}}
\maketitle

\abstract*{Here is an Abstract}
This chapter reviews how galactic inflows influence galaxy metallicity.  The goal is to
discuss predictions from theoretical models, but particular emphasis is placed on the 
insights that result from using models to interpret observations.  Even as the 
classical ``G-dwarf problem" endures in the latest round of observational 
confirmation, a rich and tantalizing 
new phenomenology of relationships between $M_*$, $Z$, SFR, and gas fraction is 
emerging both in observations and in theoretical models.  A consensus interpretation
is emerging in which star-forming galaxies do most of their growing in a quiescent 
way that balances gas inflows and gas processing, and metal dilution with enrichment.
Models that explicitly invoke this idea via equilibrium conditions can be used
to infer inflow rates from observations, while models that do not assume equilibrium
growth tend to recover it self-consistently.  Mergers are an overall subdominant 
mechanism for delivering fresh gas to galaxies, but they trigger radial flows of 
previously-accreted gas that flatten radial gas-phase metallicity gradients 
and temporarily suppress central metallicities.  Radial gradients are generically 
expected to be steep at early times and then flattened by mergers and enriched 
inflows of recycled gas at late times.  However, further theoretical work is 
required in order to understand how to interpret observations.  
Likewise, more observational work is needed in order to understand how metallicity 
gradients evolve to high redshifts.

\section{Introduction}
\label{sec:intro}

Theoretically, inflows happen.  The classical and perhaps most famous 
motivation for this view is the failure of the ``closed-box" 
model~\citep{sch63,tin80} to account for the observed paucity of 
low-metallicity G and K stars~\citep{pag75,cas04}, a discrepancy that 
persists even with the most recent measurements of stars throughout the
Milky Way disk~\citep{sch12}.  The closed-box model has other problems
such as its inability to account for the slow decline in galaxy gas
fractions~\citep{tac13} and the cosmic abundance of neutral 
hydrogen~\citep{wol05}.  Likewise, it cannot account for the weak observed
evolution in galaxy metallicities during the interval 
$z=2\rightarrow0$~\citep{erb06}, an epoch during which most of the 
present-day stellar mass formed.

More importantly, however, galaxy growth \emph{without} inflows is theoretically
incompatible with the current $\Lambda$CDM paradigm.  In this picture,
galaxies are viewed as condensations of cold baryons within dark matter halos.
The dark matter halos themselves grow via a sequence of mergers that is 
decoupled from baryon physics and straightforward to compute using either 
analytic~\citep{whi78,whi91} or numerical methods~\citep{spr05}.  As the halos
grow, they accrete gas directly from the intergalactic medium (IGM).  The
vast majority of this gas accretes in a smooth fashion; that is, it does
not arrive having previously condensed into an interstellar medium 
(ISM;~\citealt{nel13}).  Halos that are more massive than the cosmic 
Jeans mass are expected to acquire a mass of baryons that is of 
order $\frac{\Omega_b}{\Omega_M} \times M_{\rm DM}$, where $M_{\rm DM}$ is 
the halo's total mass~\citep{gne00,oka08,chr16}.  Roughly half of this 
material collapses from the halo onto the central galaxy~\citep{chr16}, 
driving further star formation.  

The expected thermal history of collapsed gas prior to its arrival in the 
central galaxy remains a topic of active study.  It was originally assumed 
that all gas is shock-heated to the virial temperature and then cools in a 
spherically-symmetric way~\citep{whi78}.  This was challenged a decade ago by 
numerical calculations, which found that much of the gas accretes 
directly onto the central galaxy without ever being heated, particularly at 
masses below $10^{12}\msun$~\citep{ker05,dek06}.  The most recent calculations
that include significantly improved hydrodynamic solvers contradict
those results, attributing the lack of shock-heating and the inefficient 
cooling of the hot gas in previous calculations to numerical 
problems~\citep{nel13}.  The new calculations indicate that the 
majority of gas at all halo masses is heated to the virial temperature 
before accreting onto the halo.  However, it does not accrete in a 
spherically-symmetric fashion as originally envisioned~\citep{whi78}. 
Instead, it tends to concentrate in coherent structures that connect to 
large-scale intergalactic medium (IGM) filaments.  The upshot is that, one 
way or another, gas readily accretes efficiently enough in $\Lambda$CDM to 
form the observed galaxy populations, with most gas arriving in the form 
of smooth inflows.

Once the gas condenses to densities of $\sim1$ atom per cm$^{-3}$,
gravitational instability triggers the formation of molecular clouds and 
eventually stars.  Feedback energy from the young stars limits the efficiency 
of star formation and regulates the ISM's structure in a number of ways.  
For our purposes, the most important of these is the generation of galactic 
outflows, which are inevitably observed wherever there is vigorous star 
formation~\citep{vei05}.  Theoretical models consistently predict that
the mass of material that is ejected is comparable to or greater than 
the mass of stars that form~\citep{mur05,mur15,chr16}.  This enriched 
material then becomes available for re-accretion after a few dynamical 
times~\citep{opp10,hen13,chr16}.  

Outflows thus give rise to two conceptually distinct gas accretion 
channels, ``Primordial Gas" and ``Recycled Gas".  Primordial gas dominates 
inflows at early times and low masses~\citep{opp10,ma16}, and it dilutes
galaxies' gas-phase metallicities.  Recycled gas becomes increasingly 
important at late times and high masses.  It is pre-enriched, and therefore 
less effective at dilution.

To summarize, in the era of \lcdm, galaxy growth driven by ongoing inflows
is \emph{unavoidable}.  The central conceit of this chapter is that 
measurements of galaxy metallicities may be used to test models of those 
inflows.  To motivate our discussion of 
how they do so, we list the observational probes 
that have been deployed:

\begin{itemize}
\item Stellar metallicity distributions;
\item The slope, normalization, and evolution of the mass-metallicity relation (MZR);
\item Third-parameter dependencies of metallicity on SFR, gas fraction, redshift, and environment; and
\item Radial metallicity gradients (chiefly of the gas).
\end{itemize}

Stellar metallicity distributions have historically been an important indicator
that inflows occur, but they are only available for the Milky Way and a handful
of its satellite galaxies~\citep{kir11}.  For this reason, we will not discuss 
them further.  Rather, we will focus on extragalactic diagnostics where larger 
samples are available.  We also note that, throughout this discussion, we will 
focus on the oxygen metallicity as it is the most widely-observed tracer of the
overall gas-phase metallicity.  In Section~\ref{sec:physProc}, we review the 
physical processes through which inflows modulate galaxy metallicities.  In 
Section~\ref{sec:galDecouple}, 
we discuss the extent to which galaxy growth tracks the host halo growth.
In Section~\ref{sec:equil}, we introduce the Equilibrium Model, which is the
simplest way for relating observables to inflows.  In Section~\ref{sec:noneq},
we discuss departures from equilibrium growth.  Finally, in 
Section~\ref{sec:summary} we summarize.

\section{Physical Processes}
\label{sec:physProc}

Gas inflows impact galaxy metallicities on a wide range of timescales and
spatial scales, and through a variety of processes that we associate either
with Mergers, Galactic Fountains, or Environment.  In this section, 
we summarize observational and theoretical progress in understanding 
these processes.

\subsection{Mergers}
\label{ssec:mergers}
% Mergers - central galaxy metallicity
Both observations and theory indicate that mergers are not the dominant way
in which fresh gas is delivered to galaxies~\citep{pap11,beh13}.  However,  
on timescales comparable to a dynamical time, mergers can cause relatively 
unenriched gas that has previously settled in a galaxy's outskirts to 
plummet into the central 100--1000 parsecs (pc), simultaneously suppressing 
the central gas metallicity and boosting the central star formation rate.  

This effect arises naturally in hydrodynamic simulations of galaxy 
mergers.~~\citet{tor12} found
that interactions suppress the nuclear metallicity of gas-poor galaxies (gas
fractions of $\leq 20\%$) by an amount that fluctuates depending on which 
merger stage the galaxies are observed in, but is typically $\approx0.07$ dex.  
Gas-rich galaxies, by contrast, can experience nuclear metallicity boosts 
during interactions although this effect is not expected in typical SDSS 
galaxies owing to their generally low gas fractions.  A similar result was 
presented
by~\citet{rup10a}, who found that the suppressed metallicities were associated
with the period between the first and second pericenter passages of a merger
event.

Does this mean that metallicities can be used to detect mergers? In fact,
the statistical signature of nuclear inflows has been identified 
observationally in galaxy pairs from the Sloan Digital Sky Survey (SDSS).  
In particular,~\citet{scu12,scu13} showed that star-forming galaxies have 
suppressed central metallicity (by 0.02--0.3 dex) and enhanced central SFR 
(by 60\%) if they are separated from a paired galaxy by a projected 
separation of $\leq\ 80\ \kpc\ h^{-1}$ (see also~\citealt{rup10b}).  It is
also possible that galaxy mass and merger mass ratio modulate this 
effect: Splitting the merging SDSS galaxies by stellar mass,~\citet{mic08} 
have argued that mergers systematically boost the metallicity of low-mass 
galaxies while suppressing the metallicity of massive ones.  The 
different behavior reflects a competition between dilution and enrichment, 
and is qualitatively expected based on the results presented 
by~\citet{rup10a} if low-mass galaxies are relatively gas-rich while 
massive ones are gas-poor.

While the ability of mergers to drive nuclear flows has been identified
both in simulations and in observations, these flows have yet to be
invoked as a rigorous constraint on galaxy evolution models.  As a first
step in this direction,~\citet{gro15} asked whether a small but robust 
population of galaxies whose metallicities were lower than expected for 
their combination of stellar mass and
star formation rate~\citep{man10} could be interpreted as ongoing mergers or merger
remnants.  They developed a simple analytical model invoking the known halo 
merger rate along with the assumption that mergers boost SFR and suppress
metallicity for a set time and showed that this model's parameters can
be tuned to yield excellent agreement with the observed outlier population.
They found that post-merger galaxies must exhibit diluted nuclear
metallicities for $1.568_{-0.027}^{+0.020}$Gyr, and the average 
dilution for mergers with mass ratios of 1:1--1:5 was 0.114 dex.  
Encouragingly, these inferences are within a factor of two of 
expectations from numerical simulations and observations.

A caveat to analyses of SDSS metallicities is that the SDSS spectroscopic
fibers subtend a diameter of 3 arcsec, hence they capture only the central 
5 kpc of galaxies at the typical redshift of SDSS galaxies ($z=0.1$).  While
this makes them ideal for studying \emph{nuclear} metallicities, it leaves
unanswered the question of how mergers effect the metallicity at larger 
radii.~~\citet{rup10a} showed that the same interaction-triggered 
flows that dilute central metallicities also flatten overall gas
metallicity gradients.  This effect has likewise been observed in ongoing 
mergers~\citep{kew10}.  Unfortunately, it is difficult to use the physics
of ongoing mergers to test galaxy formation models because results are
so sensitive to details of the merger such as gas fractions, mass ratio,
and orbital parameters~\citep{tor12}.

Fortunately, the tendency for mergers to flatten metallicity gradients
and to build up central bulges does leave an observable signature: Galaxies 
that are more massive or have larger (classical) bulges should have flatter 
gas metallicity gradients.~~\citet{fu13} have shown that this effect is strong 
observationally, and can readily be reproduced by their semi-analytical 
model (their Figure 12).  However, they did not explore which of the 
assumptions underlying their merger model are required to yield the good
agreement.  Moreover, they address only local observations, and suggest that
extending the work to higher redshifts would be a useful next step.  We will
return to this point in Section~\ref{ssec:equil:multiZone}.

\subsection{Outflows and Galactic Fountains}
\label{ssec:galFountains}

The way in which galactic inflows modulate metallicities and metallicity
gradients depends critically on galactic fountains.  Observations indicate 
that vigorous star formation is inevitably associated with outflows that 
eject gas from galaxies at a rate that is comparable to the star formation 
rate~\citep{vei05}.  Observations and models both suggest that they carry 
away most of the metals that are generated in core-collapse 
supernovae~\citep{mar02,fu13}.  The gas is expected to re-accrete on a dynamical 
timescale, particularly in massive galaxies ($\log(M_*/\msun) > 10.5$) and 
at relatively late times ($z<1$;~\citealt{opp10,hen13}).  The gas that a 
central galaxy re-accretes contains contributions both from its 
progenitor and from satellite systems, while satellite systems probably
do not accrete much gas.  Importantly, while the ejected gas has 
predominantly low angular momentum, it picks up angular momentum from 
the halo and re-accretes at generally larger radii than where it was 
ejected~\citep{bro12,chr16}.  This has a number of observational 
consequences.  

\subsubsection{Galaxy-averaged quantities}
\label{sssec:galAve}

First, re-accreted metals boost the central galaxy's metallicity with
respect to models with unenriched inflows (or, viewed from a 
different perspective, they suppress dilution).  If the metallicity
of inflowing gas varies with mass, then recycled gas must affect the
\emph{slope} of the mass-metallicity relation.  In particular, if 
low-mass galaxies accrete predominantly pristine material while 
inflows into massive galaxies are pre-enriched~\citep{ma16,bro14}, 
then this differential pre-enrichment steepens the slope of the 
mass-metallicity relation.

If pre-enrichment levels vary with redshift, then they furthermore 
contribute to the evolution of the normalization of the mass-metallicity 
relation~\citep{dav11}.  For example, if inflows are pristine at early 
times and enriched to roughly the same level as
the ISM at $z=0$, then they are more effective at diluting gas reservoirs
at high redshift, driving stronger evolution in the normalization of the
mass-metallicity relation.  The normalization of the MZR is observed to
increase~\citep{mai08,fai16}, which could be explained in this way.
The observational challenge is to disentangle this factor from other 
influences that could also evolve with redshift (or mass) such as the 
initial mass function~\citep{pvd08,dav08}.

It is instructive to compare this interpretation of how the MZR's 
normalization evolves upwards to that presented in~\citet{ma16}.   In 
the latter work, high-resolution simulations were used to show that
the relationship between gas metallicity and stellar mass fraction $f_*$ 
(that is, the stellar mass divided by the baryonic mass) is nearly that
of a closed-box throughout $z=3\rightarrow0$ when averaged over the entire 
halo rather than just the central galaxy.  In this view, growth in the
MZR's normalization tracks growth in $f_*$.  While this is a suggestive
insight, it does not directly address growth in \emph{galaxy} metallicities, 
which are far more readily observable.  It is not hard to imagine, for
example, that gas flows might hide metals within the halo
at high redshifts but then shift progressively more of them into 
observability at late times.  In other words, galaxy metallicities may 
evolve to become an increasingly unbiased probe of the 
halo metallicity, which in turn tracks closed-box expectations.  A 
more detailed study of the evolving relationship between halo-averaged 
and ISM-averaged metallicity is probably indicated.  For the present, 
however, it is clear that pre-enriched inflows have the potential to 
drive MZR evolution.

\subsubsection{Radial Metallicity Gradients}
\label{sssec:radGrad}
If inflows deposit gas that has relatively uniform metallicity 
over a range of radii as expected theoretically~\citep{bro12,chr16}, then 
they flatten radial metallicity gradients because the metallicity at any 
point is driven by the inflowing gas, washing out other influences such
as radial gradients in star formation efficiency or wind 
characteristics, or radial flows that escort low-metallicity in from the
galaxy's outskirts.  This raises the possibility of using observed 
metallicity gradients to constrain inflows.

At low redshift, it has long been known observationally that 
star-forming galaxies have slowly-declining metallicity 
gradients~\citep{zar94}, and recent analyses have confirmed these
results~\citep{fu13,car15,ho15}. \citet{fu13} used a semi-analytical
model of galaxy formation to interpret observations.  They showed that 
inflows can readily dominate metallicity gradients (their Figure 6), while 
the effect of radial flows is probably relatively weak.  They further found 
that it was necessary to assume that 80\% of all newly-formed metals 
are launched into the halo in order to match observed metallicity gradients, 
qualitatively consistent with inferences from X-ray observations of local 
outflows~\citep{mar02}.  In their model, the tendency for ejected metals
to re-accrete over a range of radii makes the baryon cycle into an 
efficient method for redistributing metals.

\citet{car15} interpreted locally-observed metallicity gradients using a 
simpler model that assumes a local equilibrium between inflows, star formation, 
and outflows (Section~\ref{ssec:equil:multiZone}).  In their model, radial 
flows are ignored, and inflows are assumed to be uniform across the disk.  
Hence the observed metallicity gradient is driven by the radial dependence 
of the mass-loading factor (that is, the ratio of the outflow to star formation
rates), with weaker outflows yielding higher metallicity 
in the center and stronger outflows suppressing
it toward the disk edge.  This model requires inflows in order to
balance ongoing enrichment~\citep{lil13}, but that does not mean that 
the data require strong inflows.  In fact, it is not even possible to use 
the~\citet{car15} model to measure inflow rates or metallicities owing 
to degeneracies.  For example, a high observed metallicity could reflect 
a high metal yield and unenriched inflows, or a low metal yield and 
highly-enriched inflows.  

The model explored by~\citet{ho15} leads to a different interpretation,
tying locally-observed metallicity gradients to gradients in gas fraction.  
These authors relax the assumption that all regions of a galaxy are in 
local enrichment equilibrium, but assume that the mass-loading factor and 
the ratio of inflow to 
star formation rates $\dot{M}_{g,\mathrm{in}}/$SFR are all constant.
In this case, metallicity decreases with radius because more diffuse
regions have higher gas fraction, implying that they are chemically 
less mature.  They also find that their models fit observations if they
assume that both inflows and outflows are weak, indicating nearly
closed-box chemical evolution for systems with low gas fraction.

The~\citet{ho15} and~\citet{car15} studies both leverage high-quality
measurements of radial trends in metallicity and gas fraction, but their
modeling efforts lead to different conclusions regarding the flow of
gas into and out of the galaxy because they invoke different 
assumptions.  Which model is more correct? More theoretical insight into
how
observables connect to the underlying physics would certainly help.
At the same time, it is to be hoped that future studies that leverage 
measurements on halo metallicities or inflow rates (from, for example, 
absorption-line campaigns) will eventually break the underlying 
degeneracies.

\citet{fu13} indicated that the evolution of metallicity gradients to
high redshift is a complementary constraint on star formation and gas
flows.  This echoes~\citet{pil12}, who analyzed how the radial and 
vertical gradients evolved in 25 cosmological simulations of Milky Way 
analogs from several groups as well as two independent analytical models.
They found that, although all models roughly reproduce the Milky Way's 
current radial gradient, some predicted dramatically steeper gradients
at earlier times while others did not.  They concluded that steep
gradients in metallicity reflect steep gradients in star formation 
efficiency and noted that strong feedback can wash out metallicity 
gradients.  However, their discussion did not consider the possible role 
of inflows in balancing enrichment.

Observationally, the high-redshift story is far from clear.  An early 
integral field study of three star-forming galaxies at $z\sim3$ uncovered 
\emph{inverted} metallicity gradients (i.e., the gas-phase metallicity 
is lower in the center;~\citealt{cre10}).  The authors interpreted their
findings as evidence that inflows deliver pristine gas preferentially
to the center of high-redshift galaxies, although in fact this is not 
expected to produce inverted gradients generically~\citep{pil12}.

Over the following years, detailed analyses of a few strongly-lensed 
high-redshift systems yielded steeply declining metallicity
gradients~\citep{yua11,jon13} as predicted in~\citet{pil12}.
These results seemed to indicate that strong inflows do not flatten 
or invert metallicity gradients at high redshifts.  

Most recently, however, a study of a large sample of unlensed systems 
indicated that metallicity gradients are flatter than in the local Universe 
or even absent at higher redshifts~\citep{wuy16}.  As noted 
in~\citet{wuy16}, it is not obvious why lensed systems should show strong 
gradients while unlensed ones do not, particularly given that the
samples overlap in stellar mass.~\citet{wuy16} consider a number of 
effects such as beam-smearing, AGN, or shocks that could artificially 
suppress the intrinsic metallicity gradient in unlensed systems, but
conclude that they would have detected strong gradients if they were 
there.  Hence while further work is needed to control biases, current 
observations support the idea that strong inflows flatten high-redshift 
metallicity gradients, as qualitatively suggested by some (but not all) 
models~\citep{bro12,fu13}.

\subsubsection{Future Work}
\label{sssec:futWork} At a technical level, the pioneering study of re-accretion
presented in~\citet{opp10} deserves to be re-visited in the context of more
recent numerical models for two reasons.  The less important of these is
that galactic outflow models have grown more realistic owing to high-resolution
simulations~\citep{mur15} as well as to increasingly detailed comparison with 
measurements~\citep{mit15}.  The more important reason is that simulations now incorporate 
significantly improved hydrodynamic solvers and dynamic range, which are critical
for resolving the complicated interaction between outflows and the circumgalactic
medium~\citep{nel15}.  Relatedly, improved cross-fertilization of insights 
regarding outflows and enriched inflows between hydrodynamic simulations and 
semi-analytical models would be helpful both for distilling insight from the
numerical models and for exploring its implications within a more flexible 
framework.

With an improved understanding of the baryon and metal cycles, a straightforward
next step would be to review the hypothesis presented in~\citet{dav11} that 
pre-enriched inflows drive the MZR's shape and evolution.  A detailed budgeting
of how galaxies distribute their baryons and metals within the ISM and halos 
would be necessary for this step, and would inform the next generation of
measurements of the CGM, which has become a very active field over the past
few years.

In the specific case of merger-induced inflows, it is not to soon to ask
whether simulations can accommodate the rich set of observations of how
interactions trigger nuclear flows driving star formation, gas dilution,
as well as AGN activity (cf. the ``Galaxy Pairs in the Sloan Digital
Sky Survey" paper series by S.\ Ellison and collaborators); this would
test the hypothesis that mergers can be identified based on their 
suppressed nuclear metallicities~\citep{gro15}.

Finally, the effective use of metallicity gradients for constraining gas 
inflows in a cosmological context will require further development both 
in models and in observations.  An improved understanding of what drives 
the metallicity gradients that are predicted by numerical simulations, a 
converged picture for the observed evolution of metallicity gradients to 
higher redshift, and the continued development of techniques for using 
observations to test models (for example, accounting for biases associated 
with sample selection and and strong-line abundance indicators) will yield 
further insight into the baryon cycle.

\subsection{Environment}
\label{ssec:environment}
At fixed stellar mass, galaxies that live in richer environments are observed
to have slightly higher metallicities~\citep{mou07,coo08}, although the effect
is weak and not always observed~\citep{hug13}.  A qualitatively similar 
relationship occurs in cosmological simulations~\citep{dav11}, suggesting that 
it may in fact be real.  If so, it could reflect systematic variation in
the basic properties of the baryon cycle.  For example, it is easy to imagine
that weaker outflows, more efficient re-accretion of previously-ejected gas, 
or enriched inflows could boost the metallicities in overdense regions.  Indeed,
analytical models have been used to argue that enriched inflows are a viable
explanation for the offset~\citep{pen14}.

In reality, it may be that none of these is the correct explanation.  A detailed 
study of galaxies in the {\sc Illustris} simulations~\citep{gen16} recently 
traced the metallicity-environment correlation to two causes.  First, at a 
given stellar mass, satellite galaxies 
(which dominate richer environments) tend to form earlier than centrals, 
draining their gas reservoirs and boosting their metallicities.  Second, 
satellites' disks tend to be truncated and more centrally-concentrated 
such that observations are weighted toward their metal-rich cores.  Combining
these effects essentially accounts for the the entire simulated offset.

\citet{gen16} notes that the CGM around satellite galaxies is more enriched than
around centrals of the same stellar mass, and in fact the offset is comparable
to offset in the galaxies' metallicities.  However, he argues that this does
not dominate the dependence of metallicity on environment because
centrals and satellites with similar stellar mass and star formation 
history have nearly the same metallicity---despite the apparently higher 
metallicity of the gas around satellites.

Future work disentangling the metallicity of inflowing and outflowing gas may
be needed in order to understand how a more enriched CGM does not drive higher
galaxy metallicities.  For the present, however, it seems that the tendency
for metallicity to increase in rich environments does not relate to inflows.

\section{Galaxy Growth and Halo Growth}
\label{sec:galDecouple}

The previous section drew attention to our imperfect understanding of 
how galaxies and halos grow together.  A specific problem related to
this question is the ``dwarf galaxy conundrum", or the suite of
observational clues that galaxies with stellar masses below 
$10^{10}\msun$ do not grow at the same specific rate as their host
halos (for example,~\citealt{wei12}).  This behavior is very difficult
for conventional galaxy evolution models to accommodate~\citep{whi15,som15}: 
generically, the strong outflows that seem required in order to bring 
the predicted stellar mass function at $z=0$ into agreement with 
observations lead to metallicities and gas fractions that are
under-predicted at low redshifts.  Conversely, adopting weaker outflows 
at low masses in order to match MZR observations yields too many stars 
in low-mass halos at $z=0$.

The problem clearly points to the need for a qualitatively new physical 
mechanism that retards gas processing in low-mass systems.  This raises
the question of whether that process operates in the IGM/CGM, or in the 
ISM.  In a creative analytic
model,~\citet{bou10} argued for a CGM-based solution, showing that 
forbidding halos less massive than $10^{11}\msun$ from cooling their 
gas onto galaxies improves agreement with measurements of star formation 
and stellar mass growth.  They unfortunately did not address
metallicities.~~\citet{lil13} also addressed this problem.  They found that,
under the assumption that enrichment and dilution balance exactly
(Section~\ref{sec:equil}), the fraction of galaxy gas that is converted to 
stars can be inferred directly from the MZR, and the resulting scaling 
matches the requirement that is implied by the stellar mass function.  In 
this argument, the dwarf galaxy conundrum can be resolved by processes 
that occur within the ISM, although inflows are required.

In order to assault the problem using a 
more comprehensive suite of observables,~\citet{whi15} implement several 
galaxy-slowing mechanisms (both galaxy-based and CGM-based) into a 
semi-analytical model.  They show that two 
changes to the fiducial model can improve agreement:  in their ``preferential
reheating" model, low-mass galaxies eject systematically more gas
in outflows per unit of stellar mass formed at higher redshift.
In their ``parking lot" model, ejected gas is not permitted to
re-accrete for a delay time that depends on halo mass (see 
also~\citealt{hen13}).  By manipulating the parameters governing these 
processes, they found improved agreement with measurements of cold gas 
fractions, specific star formation rates, and ISM metallicities

In the context of numerical simulations,~\citet{ma16} have shown
that accounting more realistically for the physical processes that
occur within the interstellar medium (ISM) tends naturally to
decouple the growth of low-mass galaxies from their host halo, 
leading to improved agreement with observations of the MZR 
evolution (see also~\citealt{hop14}).  Their findings are 
qualitatively consistent with 
the inferences that~\citet{whi15} draw from semi-analytical models, 
but it is difficult to draw robust conclusions owing to the small
number of halos that have been simulated at high resolution.  For
example, is the bottleneck that retards gas processing located in
the ISM, in the CGM, or both? Ongoing efforts to distill their 
results into scalings that can be implemented into cosmological 
simulations and semi-analytical models will generate further 
insight~\citep{dav16}.

\section{The Equilibrium Model}
\label{sec:equil}

Over the last decade, an increasing number of theoretical studies 
have arrived at the conclusion that reasonably massive star-forming
galaxies ($M_* > 10^{10}\msun$) grow in a quasi-smooth fashion whereby 
gas is accreted from the CGM and processed at roughly the same rate.  
This is a generalization of the idea that star formation self-regulates 
to match the gas accretion rate~\citep{tin78,kop99} in that it 
replaces ``star formation" with ``gas processing", where by the 
latter we mean that gas is processed either into stars \emph{or} 
outflows.  In this section, we introduce the concepts governing the
equilibrium model and discuss inquiries into when and where it 
applies.

\subsection{A Single Zone}
\label{ssec:equil:singleZone}

We begin by considering a galaxy to be a single zone
and define the growth rate of stellar mass in long-lived stars 
$\dot{M}_*$, the rate of change of the gas mass
$\dot{M}_g$, the gas and star accretion rates 
$\dot{M}_{g,\mathrm{in}}$, $\dot{M}_{*,\mathrm{in}}$, and the gas
outflow rate $\dot{M}_{g,\mathrm{out}}$.  Generically, these are
related by
\begin{equation}\label{eqn:massCons}
\dot{M}_* + \dot{M}_g = \dot{M}_{g,\mathrm{in}} - \dot{M}_{g,\mathrm{out}} + \dot{M}_{*,\mathrm{in}}
\end{equation}
Equation~\ref{eqn:massCons} is a tolerably comprehensive statement of the 
conservation of baryonic mass within the context of galaxy evolution.  
Note that star formation does not appear in this equation because it does
not change the galaxy's baryonic mass.

To derive the Equilibrium Model, we begin by assuming that accretion 
of already-formed stars in actively star-forming galaxies is negligible; 
that is, $\dot{M}_{*,\mathrm{in}} = 0$.  This implies that stellar mass 
growth owes predominantly to star formation, or $\dot{M}_*=\rm{SFR}$, which
is supported observationally~\citep{pap11,beh13}.  Next, we 
assume that the outflow rate equals the star formation rate of long-lived
stars multiplied by a slowly-varying ``mass-loading factor" $\eta$; that 
is, $\dot{M}_{g,\mathrm{out}} = \eta \dot{M}_*$.  Finally, we assume that 
the galaxy balances accretion exactly with gas processing so that 
$\dot{M}_g = 0$.  This condition cannot, of course, be exactly correct, 
as it would imply that galaxies have no gas.  However, high-resolutions 
simulations suggest that, indeed, the mass of material that is processed
and ejected by disks is comparable to the mass that is 
accreted~\citep{chr16}, justifying it as a first-order approximation.  
What is left is~\citep{fin08}:
\begin{equation}\label{eqn:eq}
\dot{M}_*(1+\eta) = \dot{M}_{g,\mathrm{in}}
\end{equation}

For clarity, we note that many authors define a total star formation rate 
SFR, compute the ``return fraction", $R$, or the fraction of mass that 
collapses into massive stars and is then returned to the ISM on a short 
timescale~\citep{tin80}, and then write down the growth of stellar
mass in long-lived stars as $\dot{M}_* = \mathrm{SFR}(1-R)$.  In this
case, we define $\eta$ as the the ratio of the outflow rate to the
total star formation rate, and Equation~\ref{eqn:eq} is unchanged.

Equation~\ref{eqn:eq} is a statement of \emph{gas processing equilibrium}.
The intuition is that, if the gas accretion rate temporarily exceeds the 
gas processing rate, then the gas density increases, boosting gas processing 
until the excess has been worked through.  Conversely, if accretion 
temporarily sputters, then the gas density declines, choking
off gas processing until new inflows are available.  Equation~\ref{eqn:eq}
assumes that this balancing occurs on a timescale that is very short.  In
doing so, it departs conceptually from the traditional view that the star 
formation rate is fundamentally determined by the gas density.  In an
equilibrium scenario, the ISM adjusts itself so as to satisfy 
Equation~\ref{eqn:eq}; the gas density merely reflects the details of
how it achieves this.

The factor of $(1+\eta)$ is what distinguishes Equation~\ref{eqn:eq} from
classical analytical models~\citep{tin78,kop99}.  To see how important it is,
consider the influences that determine a galaxy's gas-phase metallicity. At
any given time, inflows from the CGM dilute the galaxy's metallicity.  
However, the inflows trigger continuing star formation, which simultaneously
enriches the gas.  The competition between dilution and enrichment gives
rise to an \emph{Equilibrium Metallicity} 
$Z_\mathrm{eq}$~\citep{fin08,dav12}:
\begin{equation}\label{eqn:Zeq}
Z_\mathrm{eq} \equiv Z_0 + \frac{y}{(1+\eta)}
\end{equation}
Here, $y$ is the metal yield, or the mass of metals ejected by core collapse
supernovae per mass of long-lived stars formed; and $Z_0$ is the metallicity
of inflowing gas, which for convenience we may re-cast as 
$Z_0 \equiv \alpha_Z * Z$, where $\alpha_Z$ is the ratio of the metallicity 
of inflowing gas to the ISM metallicity and $Z$ is the ISM metallicity.  
If $\eta$ is redefined such that $\eta\equiv\mathrm{SFR}/\dot{M}_{g,\mathrm{out}}$ 
where SFR is the total star formation rate, then we replace
$\eta\rightarrow \eta/(1-R)$ in Equation~\ref{eqn:Zeq}.  Note that
we assume that outflows have the same metallicity as the ISM; this is 
supported by high-resolution simulations (for example, Figure 11 
of~\citet{ma16}.  

Equation~\ref{eqn:Zeq} is a statement of \emph{enrichment equilibrium}.  The
intuition is that enrichment and dilution balance each other once the galaxy's
metallicity reaches $Z_\mathrm{eq}$.  If $Z$ somehow jumps above
$Z_\mathrm{eq}$, then inflows dilute it.  If $Z$ dips, then continued star 
formation boosts it back to $Z_\mathrm{eq}$.  If the SFR responds 
instantaneously to fluctuations in the inflow rate, then departures 
from $Z_\mathrm{eq}$ are erased on a \emph{dilution time} 
$M_g/\dot{M}_{g\mathrm{,in}}$, or 
the timescale for inflows to replenish the ISM completely.

Equations~\ref{eqn:eq}--\ref{eqn:Zeq} involve two parameters, $\eta$ and 
$Z_0$.  For galaxies that are massive enough to be in equilibrium, the
slope and redshift-dependence of the mass-metallicity relation may be
expressed as evolution in these parameters.

\begin{figure}[h]
%\sidecaption
\includegraphics[scale=.6]{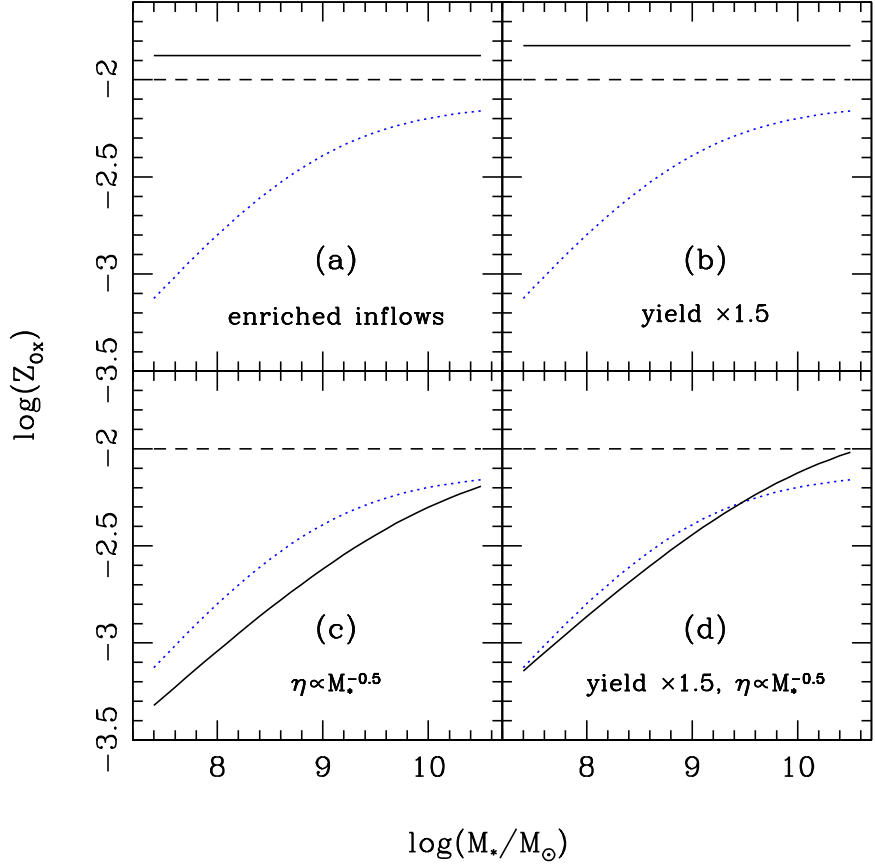}
\caption{Illustration of the parameters that control metallicity in
Equation~\ref{eqn:Zeq}.  In each panel, the black dashed line indicates 
the same ``base model" while the blue dashed curve is the observed 
trend~\citep{and13}. The bottom-right panel shows that Equation~\ref{eqn:Zeq}
can be tuned to achieve good agreement with observations.}
\label{fig:Zeq}
\end{figure}

In order to illustrate how powerful Equation~\ref{eqn:Zeq} is, we use it
to interpret the $z=0$ mass-metallicity relation (where by ``metallicity" 
we mean the oxygen mass fraction $Z_\mathrm{ox}$).  In each panel of 
Figure~\ref{fig:Zeq}, the black 
dashed line indicates $Z$ under the assumption of pristine inflows 
($Z_0 = 0$), an oxygen yield of $y=0.01$, and no outflows ($\eta = 0$).  
Meanwhile, the blue dotted curve is the observed trend at 
$z\sim0.1$~\citep{and13},
where we convert from units of $12 + \log(\mathrm{O/H})$ to oxygen mass 
fraction assuming a hydrogen mass fraction of 0.76.  

With no outflows ($\eta=0$), metallicity has no dependence on mass as
shown by the dashed black segment in each panel of Figure~\ref{fig:Zeq}.  
Increasing the inflow metallicity $Z_0$ from 0 to 0.00333 boosts $Z$ at all 
masses (panel a).  Likewise, increasing the yield $y$ from 0.01 to 0.015 
increases $Z$ at all masses (panel b).  Assuming that the outflow rate $\eta$ 
varies with stellar mass as 
$(10^{10}\msun/M_*)^{0.5}$ immediately yields a strong dependence of $Z$ 
on $M_*$ (panel c).  Finally, restoring the assumption of pristine inflows 
and tuning both the yield and the outflows yields a plausible level of 
agreement with observations (panel d).  We do not use a rigorous approach
to tune our parameters because our goal is merely to illustrate its 
potential as a conceptual framework. However, the model has been
generalized and tuned elsewhere, yielding promising insight into the 
nature of the baryon cycle~\citep{mit15,mit16}; we will discuss these 
efforts shortly.

This exercise shows that the MZR can easily be interpreted within the 
equilibrium scenario (Equations~\ref{eqn:eq}--\ref{eqn:Zeq}).  
However, the interpretation's value is limited by model's simplicity.
For example, $\eta$ and $Z_0$ could vary with redshift, and $Z_0$ could
additionally vary with mass.  For improved realism, further insight from
numerical models is required.

\begin{figure}[h]
%\sidecaption
\includegraphics[scale=.6]{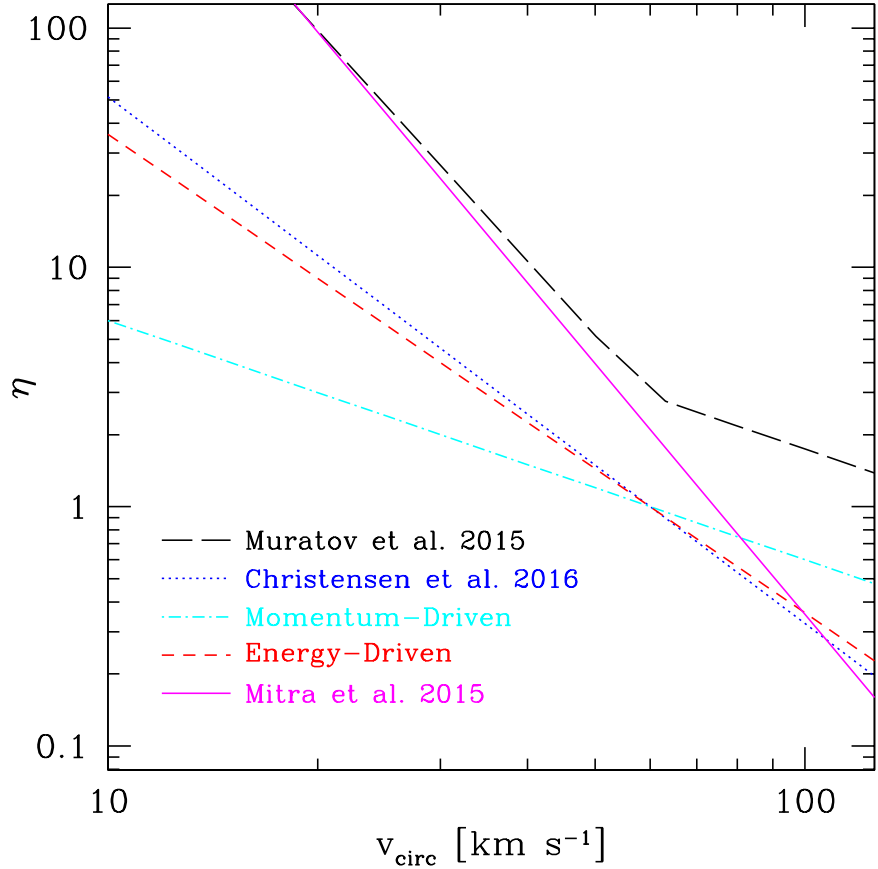}
\caption{Recent models for the ratio $\eta$ of the gas outflow rate to the
star formation rate (a.k.a. the mass-loading factor).  The models
of Christensen \emph{et al.} 2016 and Muratov \emph{et al.} 2015 are
high-resolution ab-initio simulations.  The Mitra \emph{et al.} 2015
curve results from fitting an equilibrium model to observations.  The
momentum-driven and energy-driven scalings are arbitrarily normalized
to $\eta = 1$ at $60\kms$.}
\label{fig:eta}
\end{figure}

$\eta$ is now relatively well-studied theoretically, with a consensus that it 
decreases with increasing mass and exceeds 1 in halos whose circular velocity 
$v_\mathrm{circ}$ lies below $60\kms$\citep{mur15,chr16}.  In detail, 
there is an ongoing debate as to whether it follows expectations from 
momentum-driven or energy-driven outflows.
Briefly, if the characteristic speed of galactic outflows
is the halo circular velocity $v_\mathrm{circ}$ and if star formation
deposits a constant amount of energy in outflows per unit of stellar mass
formed, then $\eta$ varies as $v_\mathrm{circ}^{-2}$.  Alternatively,
if star formation deposits a constant amount of momentum in outflows per 
unit of stellar mass formed, 
then $\eta\propto v_\mathrm{circ}^{-1}$~\citep{mur05}.

We show three recent estimates for $\eta$ in Figure~\ref{fig:eta}.
\citet{mur15} use high-resolution numerical simulations to find that 
$v_\mathrm{circ}^{-1.0}$ for $v_\mathrm{circ}>60\kms$\ and 
$v_\mathrm{circ}^{-3.2}$ at lower masses.  By contrast,~\citet{chr16} 
study a complementary suite of high-resolution simulations and
derive $\eta \propto v_\mathrm{circ}^{-2.2}$ at all masses.  These
simulations model star formation and feedback in different ways,
hence it is telling that they both predict $\eta>1$ for 
$v_\mathrm{circ}<60\kms$\ and $\eta > 4$ for 
$v_\mathrm{circ} = 30\kms$, which is the lowest halo mass that can 
efficiently accrete baryons from a photoheated IGM~\citep{gne00,oka08}.
The generally higher values of $\eta$ found by~\citet{mur15} probably
reflect additional feedback sources such as radiation pressure that
are not considered in the~\citet{chr16} model.

\citet{mit15} take a different approach and use a Bayesian Monte Carlo
Markov Chain approach to infer $\eta$ (among other things) directly from 
observations.  In particular, they assume that galaxies obey 
Equations~\ref{eqn:eq}--\ref{eqn:Zeq} at all times and then infer the 
parameters that are required in order to match 
simultaneously the observed stellar mass - halo mass relation, the MZR, and
the stellar mass - star formation rate relation.  They find that 
$\eta\propto M_h^{-1.16}\propto v_\mathrm{circ}^{-3.48}$, slightly steeper
than the dependencies obtained from numerical simulations.  When $\eta$
is expressed as a function of stellar mass, however, the differences shrink:
\citet{mur15} find $\eta\propto M_*^{-0.35}$ while~\citet{mit15} find 
$\eta\propto M_*^{-0.5}$ for $M_*<10^{10}\msun$ (note that this is the same
scaling as in panel d of Figure~\ref{fig:Zeq}).

Given that $\eta$ decreases with mass both in ab-initio calculations
and in efforts to model observations, it assuredly drives the MZR's slope: As 
shown in Figure~\ref{fig:Zeq}, a large $\eta$ removes more enriched gas, with 
the result that dilution by inflows is more efficient at suppressing the 
gas metallicity.  This in turn leads to a positive dependence of $Z$ on 
mass.

The way in which enriched inflows (that is, $Z_0>0$) affect the MZR has 
received considerably less study than $\eta$ but is probably just as 
important.~\citet{dav11} showed that, on average, $\alpha_Z$ increases 
with time, driving or at least contributing to the observed growth in the 
MZR's normalization.  The 
analysis of metal flows in high-resolution simulations presented 
by~\citet{ma16} qualitatively supports this view.  Furthermore, it suggests 
that inflows into massive halos (those that grow to $10^{11}\msun$ by $z=0$) 
have roughly one third the ISM's metallicity at all times,
or $\alpha_Z \sim \frac{1}{3}$.  Meanwhile, the $\alpha_Z$ in low-mass
halos ($10^{10}\msun$ at $z=0$) falls from $\sim\frac{1}{3}$ at early 
times to a few percent at $z=0$. 

In short, both $\eta$ and $Z_0$ probably vary in a way that drives an MZR 
to increase with mass and time.  The connection between this 
\emph{differential dilution} and the MZR depends critically on the 
equilibrium model.  This discussion then raises several points
for future research.

\emph{Which galaxies are in equilibrium?} As discussed in 
Section~\ref{sec:galDecouple}, observations suggest that low-mass halos 
depart from equilibrium expectations in the sense that they do not cool 
their gas onto galaxies at the same rate as they accrete it from the IGM.  
Is there a similar mass cutoff below which galaxies do not convert their 
stars into gas at the same rate as they accrete it from the CGM? Does it
vary with redshift? If so, then one expects the MZR normalization to
plummet above the redshift at which galaxies first achieved equilibrium.
Indeed, observations have suggested that the MZR drops rapidly in normalization
from $z=2$--3~\citep{mai08} and then flattens (that is, the mass dependence
weakens) to $z=5$~\citep{fai16}.  This behavior could indicate that $z\geq3$ is
the ``gas accumulation" epoch, when metal and dust production were 
swamped by inflows and $\eta$ and $\alpha_Z$ did not yet govern metallicities.  
A detailed analysis of what is expected from hydrodynamic simulations in this
regard would be illuminating.

\emph{What is equilibrium?}
For completeness, we note that, while some works define gas processing 
equilibrium as the condition that a galaxy's gas \emph{mass} be 
constant~\citet{fin08,dav11,dav12}, it may be equally defensible to propose
that a galaxy's gas \emph{fraction} [defined as $\mu \equiv M_g/M_*$ or $M_g/(M_g + M_*)$] 
is constant~\citep{fel13}.  If true, then the equilibrium metallicity is 
slightly different from Equation~\ref{eqn:Zeq} (i.e., set $\dot{\mu}\equiv0$ 
in Equation~\ref{eqn:Zeq_lilly2013}).  It would be interesting to ask which of 
these conditions more accurately describes galaxy growth in cosmological 
simulations or semi-analytical models.

\emph{Can we close the loop?} Once the mass and redshift range has been
identified where galaxies are expected to be in equilibrium, it would be 
interesting to ask whether Equations~\ref{eqn:eq}--\ref{eqn:Zeq} really
do govern the observed evolution within simulations and models.  
Conceptually, this would be accomplished in several (rather involved) 
steps:
\begin{itemize}
\item Identify which galaxies obey Equation~\ref{eqn:eq} or an analogous 
statement of gas processing equilibrium such as $\dot{\mu} = 0$;
\item Measure $\eta(M_*,z)$, $Z_0(M_*,z)$ and the MZR and its evolution; 
\item Ask whether Equation~\ref{eqn:Zeq} or an analogous statement of
enrichment equilibrium correctly describes the predicted MZR evolution.
\end{itemize}
A more complete conceptual understanding of what governs the MZR in
hydrodynamic simulations or semi-analytical models would inform the 
interpretation of existing and upcoming observations.

\subsection{Multi-zone models}
\label{ssec:equil:multiZone}

If the equilibrium model describes galaxies as a whole, then it is natural 
to ask whether it also describes individual regions within the galaxy.  
There is a lower limit of course; we would not apply the equilibrium model 
on spatial scales comparable to a molecular cloud (10--100 pc) where metals 
diffuse or are spread by individual supernovae.  But what about a region of 
characteristic size $\sim1 \kpc$?  To be specific, we may ask two questions:

\begin{enumerate}
\item What is the smallest region within a galaxy 
whose gas reservoir obeys Equations~\ref{eqn:eq}--\ref{eqn:Zeq}? This is 
important because the relative roles of inflows and outflows are more
difficult to assess (or even discuss) in systems that are not in 
equilibrium.
\item How does the balance between radial flows, outflows, and 
inflows vary within a star-forming galaxy?
\end{enumerate}

Within the context of an analytical model, ~\citet{fu13} found
that radial flows do not impact the overall metallicity gradient while 
inflows do.  They did not study the extent to which individual annuli
were in equilibrium, and the assumptions that were necessarily to make
their model tractable were strong, motivating further insight from 
numerical models.

A year later, a different group took the bait and carried out a detailed 
study of high-resolution, one-dimensional simulations of star-forming 
disks~\citep{for14a}.  They found that, indeed, individual annuli do obey
Equation~\ref{eqn:eq} locally.  In detail, the way in which 
cosmological inflows, star formation, galactic outflows, and radial gas 
flows (triggered by gravitational instability-driven torques that occur
wherever gas densities are high) balance varies with mass and with 
redshift.  Annuli at
large radii have low densities and low star formation rates, hence they
do not efficiently process newly-accreted gas or transport it to smaller 
radii.  Moving inwards, gas densities increase to the point that star
formation and radial transport of gas balance accretion.  Near a galaxy's
center, star formation and outflows are so efficient that they can only 
be balanced by radial transport of gas from larger radii; cosmological 
inflows are a subdominant gas source here.  The slowly-declining cosmological 
infall rate causes this balance to evolve in such a way that the region 
where gas is gravitationally unstable migrates outwards with time.  By 
$z=0$ it operates only in galaxies' outskirts, if at all.

The finding that high-redshift galaxies obey Equation~\ref{eqn:eq} 
locally as well as globally certainly raises the hope that 
Equation~\ref{eqn:Zeq} also applies, but~\citet{for14a} unfortunately
do not comment on the predicted metallicity gradients or their evolution.  
However, simple applications of this idea have been able to accommodate 
locally-observed radial trends fairly well~\citep{car15}. 

It would be interesting to compare high-resolution one-dimensional
simulations~\citep{for14a} with lower-resolution cosmological 
simulations~\citep{pil12} in order to assess (1) Whether cosmological 
simulations resolve gravitational instabilities and their impact; 
(2) Whether there are effects in three-dimensional simulations (such
as bars) that modulate gas flows and that cannot be captured in 
one-dimensional simulations; and most broadly (3) Whether the 
cosmological simulations predict the same level of equilibrium as the
one-dimensional simulations.  Likewise, it would be useful to
distill insights from the~\citet{for14a} study (such as the the
radial flow rate) into a form that can be used to improve the realism 
of semi-analytical models.

We conclude from these studies that the expected signature of 
cosmological inflows in galaxies' metallicity gradients remains 
unclear theoretically.  There is room both for an 
improved understanding of observed metallicity gradients 
(Section~\ref{ssec:galFountains}), and of what they are telling 
us about inflows and gas processing.

\section{Extensions to the Equilibrium Model}
\label{sec:noneq}

Observations indicate that there is more to galaxy metallicities than 
equilibrium.  For example, at a fixed stellar mass, galaxies show finite 
scatter in gas-phase metallicity~\citep{tre04,zah12,guo16}, and in star 
formation 
rate~\citep{whi12}.  Do these observations reflect intrinsic scatter in 
$\eta$ or $Z_0$, or do they indicate departures from equilibrium? 

The observed level of scatter is intuitively consistent with the idea
that star formation is at least slightly irregular or ``bursty" at any 
mass, which could be associated with departures from equilibrium.  For
example, if star formation rate is driven by stochastic 
processes~\citep{ger80} or if the ISM responds slowly to inflow 
fluctuations, then the galaxy would not be perfectly in 
equilibrium~\citep{for14b}; this is in fact expected for dwarf 
galaxies~\citep{lil13,hop14,wis14}.  Alternatively, the burstiness could
owe to irregular inflows, which are not themselves completely smooth.
Indeed, it has been shown that, by modeling the bursty inflow rates 
realistically but retaining the assumption that galaxies are in perfect 
equilibrium with those bursty inflows 
(which is theoretically supported~\citealt{rat16}), 
the observed level of scatter can 
be accounted for~\citep{mit16,dut10}.  This means that the observed scatter 
need not reflect truly out-of-equilibrium behavior.  

The most important handle on the nature of the observed
scatter is the identification of \emph{correlated departures} from
equilibrium, in particular departures from mean-trend behavior in $Z$, 
SFR, and gas fraction at constant $M_*$, as quantified by mean scaling 
relations.  These have prompted a range of generalizations to the 
Equilibrium Model.  Before delving into the models, however, we 
review the observations.

%\subsection{The $M_*$ - $Z$ - SFR (or $f_g$ Relation: Observations and Intuition}
\subsection{The $M_*$-$Z$-SFR Relation: Observations and Intuition} 
\label{ssec:MZSFR}

\citet{ell08} first pointed out that the SDSS MZR has a third-parameter 
dependence on SFR in the sense that, at a fixed $M_*$, galaxies with higher 
SFR have lower $Z$.  Subsequently, other authors have found qualitatively 
similar results at a variety of 
redshifts~\citep{man10,lar10,cre12,hun12,henry13,sto13,cul14,mai14,nak14,del15,salim15,kac16}.  
There have been suggestions that the SFR-dependence reflects observational 
systematics or disappears at high redshift~\citep{san13,ste14,san15}.
However, detailed analyses of possible biases have shown that it persists 
even if a variety of effects is taken into account both in the SDSS 
sample~\citep{and13,salim14,tel16} and at high redshift~\citep{salim15}.  
Moreover, early suggestions that the sign of the effect has a mass 
dependence in the sense that massive galaxies with enhanced SFR have 
enhanced metallicity~\citep{yat12} may simply reflect tricky selection 
biases~\citep{salim14,salim15}.  Initial results suggesting that the 
$M_*$-$Z$-SFR relation is redshift-independent gave rise to the idea that 
the $M_*$-$Z$-SFR relation is ``fundamental"~\citep{man10}, which would be
powerful if true.  However, more recent results indicate that, while the \emph{sense}
of the SFR dependence is unchanged throughout $z=$0--2.3, its \emph{strength}
evolves~\citep{bro16,gra16}, particularly at high stellar masses 
($10^{11}\msun$;~\citealt{salim15}).

While the dependence of metallicity on SFR was identified first, it has
recently been argued that the dependence on gas fraction is just as
strong, if not stronger~\citep{hug13,bot13,lar13,bot16}. This result has
also been found in cosmological simulations~\citep{lag16}.  These findings
indicate that SFR perturbations really just serve as a proxy for gas supply 
perturbations.  In other words, the $M_*$-$Z$-SFR relation is driven by
bursty inflows rather than bursty star formation~\citep{dut10,mit16}.  
SFR is much easier to measure for large samples, however, and will 
probably continue to be used as the \emph{de facto} third parameter 
for some time.

It thus appears that the $M_*$-$Z$-SFR relation is real, if perhaps a bit 
elusive.  Theoretical explanations generally invoke a picture along the 
following lines: suppose that a galaxy is growing in a quiescent way that 
obeys Equations~\ref{eqn:eq}--\ref{eqn:Zeq}.  If, for some reason, the gas
accretion rate increases, two things will happen.  First, the gas mass will
grow, likely suppressing the galaxy's metallicity.  Second, the SFR
will increase owing to the boosted gas density.  Finally, the 
perturbations to $Z$ and SFR subside once the extra gas is processed and the 
metallicity is diluted by further inflows~\citep{dal07}.

Of course, one
could also imagine the opposite effect: Dips in the gas accretion rate 
could lead to suppressed SFR and enhanced metallicity (because the metals
generated by ongoing star formation would be less diluted by inflows).
The problem with this is that, whereas accretion spikes can increase the 
size of the gas reservoir and boost SFR on a disk dynamical time 
($\sim100 \Myr$), accretion dips can only suppress the SFR on the timescale 
given by the gas mass divided by the SFR.  This is, to within a factor of 
a few, the same as the halo specific growth rate, or 1--10 Gyr (see, for
example, Figure 3 of~\citealt{lil13}).  It is perhaps for this reason that,
observationally, the $M_*$-$Z$-SFR dependence is driven by systems with 
enhanced SFR~\citep{salim14,tel16}, and evidence for short-term accretion 
dips is hard to come by.

A variety of theoretical approaches have been used to study the scatter in
the $M_*$-$Z$-SFR relation and its possible use as a probe of gas inflows.
We now summarize key insights gleaned from these efforts.

\subsection{The $M_*$-$Z$-SFR Relation: Equilibrium Treatments}
\label{ssec:noneq-MZSFR1}

\subsubsection{The Power of Assuming $\dot{Z} = \dot{M}_g = 0$: Dav{\'e} et al. 2012}
Inspired by cosmological hydrodynamic simulations,~\citet{dav12} devised
an equilibrium model that relates galaxy growth to host halo growth under
the condition that galaxies obey Equations~\ref{eqn:eq}--\ref{eqn:Zeq}.
In order to connect halo growth with galaxy growth, they
parameterized the impact of photoionization heating, quenching by accretion 
shocks, and re-accretion of previously ejected gas.  They also studied
the condition for galaxies to be in gas processing equilibrium, finding 
that there is a characteristic redshift above which galaxies cannot process
their gas as rapidly as they accrete it.  The precise redshift depends both 
on halo mass and on the outflow model because stronger outflows
lead galaxies to reach equilibrium sooner.  For momentum-driven outflow
scalings, $10^{12}\msun$ halos achieve equilibrium at $z\sim5$ while
galaxies in lower-mass halos ($\leq10^{11}\msun$) are always in 
equilibrium (because their outflows are strong).  Hence star-forming 
galaxies are in or near equilibrium for most of cosmic time.  

Equations~\ref{eqn:eq}--\ref{eqn:Zeq} are too simple to permit a 
third-parameter dependence of $Z$ on SFR, but~\citet{dav12} did note that
such a dependence arises naturally in hydrodynamic simulations and can
readily be attributed to weak violations of the condition 
$\dot{M}_g = 0$.

The~\citet{dav12} picture has since been generalized into an eight-parameter 
model and tuned to match the observed relationships between $M_*$, $Z$, SFR, 
and halo mass as well as their redshift dependence and 
scatter~\citep{mit15,mit16}.  The resulting fits are excellent, supporting
the model's realism.  Notably, three of the eight parameters describe the
rate at which ejected gas re-accretes and two describe ``quenching", or
the suppression of inflows into galaxies hosted by massive halos.  Such
quenching could be associated, for example, with active galactic nuclei
or virial shocks.  These studies show forcefully that, within a suitable 
framework, it is possible to use observations of $Z$ and SFR to constrain 
the efficiency of inflows.

\subsubsection{Relaxing Gas Processing Equilibrium: Lilly et al. 2013}
Initial efforts to interpret ensemble measures of galaxy evolution within 
an equilibrium framework assumed that the galaxy's gas reservoir does not 
grow~\citep{fin08,dav11,dav12}.  However, it was soon realized that relaxing 
this assumption and allowing the gas fraction to fluctuate naturally 
gives rise to a $M_*$-$Z$-SFR relation that qualitatively resembles
observations~\citep{lil13}.  Re-deriving the condition for enrichment
equilibrium while relaxing the assumption of gas processing equilibrium
leads to a generalization of Equation~\ref{eqn:Zeq}: 
\begin{equation}\label{eqn:Zeq_lilly2013}
Z_\mathrm{eq} \equiv Z_0 + \frac{y}{(1+\eta) + \epsilon^{-1} [ \frac{\mathrm{SFR}}{M_*} + (1-R)^{-1}\frac{d\ln\mu}{dt}]}
\end{equation}
Here, $\epsilon^{-1}$ is the star formation timescale and $\mu \equiv M_g/M_*$
is the ratio of gas mass to stellar mass.  Equation~\ref{eqn:Zeq_lilly2013}
beautifully connects  $Z$, SFR, and $M_*$.  If the timescale $\epsilon^{-1}$ 
for a galaxy to run through its gas reservoir is very short, then $Z$ should 
not correlate with SFR because the ISM recovers quickly from perturbations 
(by processing gas efficiently).  In this case, the gas mass is constant and
Equation~\ref{eqn:Zeq_lilly2013} reduces to Equation~\ref{eqn:Zeq}.
On the other hand, if star formation is less efficient ($\epsilon^{-1}$ is 
longer), then the ISM can be out of gas processing equilibrium for some time.  
In this case, a high specific star formation rate $\mathrm{SFR}/M_*$ is 
associated with a low metallicity as observed, irrespective of the 
relationship between the inflow and gas processing rates.

After presenting this model,~\citet{lil13} used it to draw a number
of inferences from observations.  In particular:
\begin{enumerate}
\item If the model is tuned to match the observed MZR at $z=0$, then 
it immediately predicts the strong observed dependence of stellar 
mass on halo mass~\citep{mos10}, suggesting that the latter trend 
may be governed entirely by the details of feedback and gas 
processing that occur deep within the ISM.  This does not rule out 
the possible importance of suppression or feedback processes that 
occur in the CGM, but it indicates that they are not required.
\item The model can readily be tuned to match observations of the
$M_*$-$Z$-SFR dependence, and it predicts that this relation does
not evolve with redshift unless the parameters governing gas
processing (i.e., $\eta$, $Z_0$, and $\epsilon)$ also evolve.  By
contrast, it predicts that the MZR's normalization increases
with time because high-redshift galaxies have systematically 
higher specific star formation rates.
\end{enumerate}

The gas regulator model contains a number of approximations, chief
among them the assumption of enrichment equilibrium.  It is naturally 
of interest to ask whether the model describes what happens within a less 
approximate framework.~~\citet{pip14} tested the~\citet{lil13} model using 
exact analytical expressions for the evolution of gas-phase metallicity, gas 
mass, and star formation rate, finding excellent agreement across a wide 
range of potential growth histories.  
This suggests that the gas regulator model is a powerful generalization 
of the equilibrium model introduced by~\citet{dav12}.  Not surprisingly,
it is now being tested and applied in a wide variety of 
contexts (for example,~\citealt{bir14,dek14,bou15,wu16}).

To date, there has been no effort to determine whether the~\citet{lil13} 
model describes what happens within three-dimensional hydrodynamic 
simulations or semi-analytical models; this may be a promising avenue 
for future research.

%\subsection{The $M_*$ - $Z$ - SFR (or $f_g$) Relation: Non-equilibrium Treatments}
\subsection{The $M_*$-$Z$-SFR Relation: Non-equilibrium Treatments}
\label{ssec:noneq-MZSFR2}

\subsubsection{An Analytical Model Without Equilibrium: Dayal et al. 2013}

Can the observed $M_*$-$Z$-SFR relation can be explained without any 
explicit assumption of equilibrium? In a purely analytical 
study,~\citet{day13} presented a minimal model for
galaxy growth in which the evolution of the stellar, gas, and metal
reservoirs depends only on the unknown gas accretion
and outflow rates.  They assume that outflow and infall rates
depend only on stellar mass, that the metal yield $y$ is constant,
and that inflows are pristine.  They then show that they can fit the
model's governing parameters using the the observed $M_*$-$Z$-SFR 
relation~\citep{man10} to achieve good agreement.  

In common with~\citet{lil13}, this model predicts no evolution
in the overall $M_*$-$Z$-SFR relation for a given set of parameters,
but it has the advantage that it assumes neither gas processing
or enrichment equilibrium.  Nonetheless, a mass-dependent balance 
between inflows and outflows emerges as a a central prediction.
Massive galaxies have weak winds and rapidly achieve enrichment 
equilibrium (similar to Equation~\ref{eqn:Zeq}, although they do 
not impose this).  By contrast, low-mass galaxies lose most of 
their metals through much stronger outflows and are hence more 
susceptible to dilution.  

In fact, they show that, subject to their 
modeling assumptions, the infall rate and its mass dependence 
can be inferred directly from the observed $M_*$-$Z$-SFR relation.
Intriguingly, the inferred outflow rate $\eta$ varies
with mass in a way that follows expectations from momentum-driven 
outflow models~\citep{mur05}.  The model can also accommodate
high-redshift observations, with the result that gas fractions and
inflow rates were previously higher while outflow efficiencies were
essentially the same~\citep{hun16}.

Their model also naturally reproduces the observed relation between 
oxygen abundance and neutral gas fraction~\citep{hug13}.  This finding
reinforces the tight connection between metallicity and gas fraction 
that emerges in any plausible model of galaxy growth.  It would be
interesting to ask whether re-writing the~\citet{day13} model to match
the observed $M_*$-SFR-gas fraction relation automatically recovers
the $Z$ dependence.  Likewise, it would be interesting to quantify 
precisely how close the galaxies are to enrichment and gas processing 
equilibrium in the~\citet{day13} model once the inflow and outflow 
parameters are tuned to match observations.

\subsubsection{Stochastic Accretion Histories and Scatter: Forbes \emph{et al.} 2014}
Observationally, there is residual scatter in $Z$ even after the SFR 
dependence is removed~\citep{salim14,salim15}.  Can this scatter be used 
as a complementary probe of 
inflows and gas processing? This question is taken up by~\citet{for14b},
who propose that, even if individual galaxies are not in equilibrium, 
they may be close enough to it that ensemble statistics such as the slope 
of the $Z$-SFR relation are in equilibrium.  They explore this idea by 
modeling galaxy 
growth as a connected sequence of intervals whose constant duration is 
given by the ``coherence time" $t_\mathrm{coherence}$.  At the start 
of each interval, a new accretion rate is drawn from a log-normal 
distribution with fixed median and scatter.  The galaxy then uses 
star formation and outflows to respond to the new accretion rate on 
a ``loss timescale" $t_\mathrm{loss}$, which incorporates gas losses to 
both outflows and star formation.  If $t_\mathrm{loss}$ is short
compared to other timescales, then galaxies are always in equilibrium.
Conversely, if $t_\mathrm{loss}$ is long, then they retain a ``memory" 
of previous accretion episodes.

The model's parameters can be tuned to yield qualitative agreement with
the observed $M_*$-$Z$-SFR relations.  This has two important implications:
\begin{enumerate}
\item Galaxy metallicities, when combined with measures of $M_*$ and SFR,
may probe both the mean and the distribution in gas accretion rates;
\item As in~\citet{day13}, observations do not require galaxies to obey 
either of the equilibrium relations (Equations~\ref{eqn:eq}--\ref{eqn:Zeq}) 
explicitly, although in practice they may be nearly in equilibrium.
\end{enumerate}

The price for relaxing the assumption of equilibrium is that this model
is more complicated than the~\citet{lil13} or \citet{dav12} models.
It would therefore be useful to ask how close galaxies are to equilibrium 
when this model is tuned to reproduce observations.  Moreover, as the
authors point out, the~\citet{for14b} model for inflow fluctuations is 
extremely simple, invoking only a single coherence timescale.  In
reality, inflows fluctuate on a range of timescales.  It would therefore
be useful to generalize this aspect of the model for improved realism.

\subsubsection{Cosmological Hydrodynamic Simulations}
Cosmological hydrodynamic simulations and semi-analytical models make
fewer assumptions regarding inflows and gas processing than analytical 
models~\citep{som15}, and are hence a particularly realistic 
framework for studying the $M_*$-$Z$-SFR relation.  Reassuringly, 
they inevitably predict relations that qualitatively resemble 
observations~\citep{dav11,yat12,obr14,der15,kac16,lag16,cou16}.  

In order to understand this prediction better,~\citet{lag16} apply
a principal component analysis to the predicted ensemble distribution of
global galaxy attributes in the {\sc Eagle} simulations~\citep{sch15}.  
In agreement
with previous work, they identify a particularly thin plane in the space 
of $M_*$, SFR, and neutral+molecular gas fraction where simulated galaxies 
tend to live.  Moreover, they show that this plane is observed.  They
interpret its existence as an indication that galaxies live in a 
slowly-evolving balance between gas accretion and gas processing reminiscent 
of Equation~\ref{eqn:eq}.  Interestingly, the predicted scatter about this 
plane tightens if stellar feedback assumed in the model is stronger.  
This could readily reflect the fact that fresh inflows drive galaxies 
back to their equilibrium state more quickly if the gas fraction is 
lower.

Just as the statistics of the {\sc EAGLE} galaxies imply a balance between 
gas inflows and processing, they also imply a balance 
between dilution and enrichment: the predicted metallicity is correlated
with the gas fraction in the observed sense, namely that galaxies with
high gas fractions have systematically low metallicities.  Interestingly,
however, it is found that $Z$ contributes less to the overall
variation in galaxy properties than $M_*$, SFR, and gas fraction: whereas
the latter three attributes dominate the first principal component, $Z$ 
only enters into the second.  This qualitatively supports observational 
suggestions that gas fraction is more fundamental than 
metallicity~\citep{hug13,bot13,lar13,bot16}.  Further theoretical work 
is needed in order to understand why gas fraction is more fundamental.

Unlike recent high-resolution simulations of individual 
galaxies~\citep{chr16,ma16}, the study presented in~\citet{lag16} stops 
short of directly measuring the flow of gas and metals through the
simulated galaxies because it is based on a principal-component analysis, 
which leverages the statistical advantage that lower-resolution 
cosmological simulations have over high-resolution ``zoom" studies.
This means that they do not ask whether galaxies keep gas mass or gas 
fraction constant, or how $Z_0$ varies with mass and redshift.  It would 
therefore be interesting to merge the two approaches by quantifying the 
expected flow of gas and metals in cosmological simulations in order 
to understand in more detail how they give rise to the $M_*$-$Z$-SFR-gas 
fraction relation.

\subsubsection{An Analytical Model Without Inflows: Magrini \emph{et al.} 2012} 
Thus far, all of the models for the $M_*$-$Z$-SFR relation that we have 
discussed invoke inflows, giving the impression that cosmological 
inflows of pristine gas are an unavoidable inference.  A different 
perspective is offered by~\citet{mag12}, who construct analytic,
one-zone, closed-box evolutionary models that qualitatively match 
the observed $M_*$-$Z$-SFR relation.  The key 
ingredient in their model is a distinction between ``active" and
``passive" modes of star formation: active-mode galaxies have
high-density molecular clouds whose collisions trigger vigorous
star formation while the clouds in passive-mode 
galaxies are more diffuse.  When they confront their model with
observations, they find that most high-redshift galaxies and the
occasional, unusually active low-redshift ones can be identified
with the active mode while quiescent systems can be identified
with the passive mode.

While it is interesting that the model in~\citet{mag12} contains observations 
within its accessible parameter space, the discussion leaves several basic 
questions unaddressed.  First, the model does not predict the 
distribution of active and passive-mode systems or its evolution.
Hydrodynamic simulations attribute the prevalence of active-mode 
star-formers at high redshift and the scaling relations that they obey to 
high gas accretion rates and the detailed way in which galaxies achieve
an equilibrium with those inflows~\citep{for14a,lag16}. By contrast 
(and similarly to~\citealt{day13}),
\citet{mag12} attribute the observed scaling
relationships to unexplained variations in initial conditions, evolutionary 
state, and ISM properties. 
Second, the lack of inflows means that the~\citet{mag12} predicts stellar 
metallicity distributions that conflict with local observations 
(Section~\ref{sec:intro}).  Future work that includes inflows following
cosmological expectations may reconcile this model with simulations.

\section{Summary}
\label{sec:summary}
Over the past 20 years, a marked dichotomy has emerged that distinguishes
between how dark matter halos and galaxies grow.  While halos 
do much or most of their growing by merging with other halos,  star-forming 
galaxies with stellar masses below $\sim10^{11}\msun$ acquire most of their 
baryons through relatively smooth gas flows.  Theoretical models predict 
that these inflows leave a number of observable signatures in galaxies' 
gas-phase metallicities.  Mergers leave a statistically-detectable imprint 
both on galaxies' central gas-phase metallicities and on their radial 
metallicity gradients, but they are not the primary mechanism for 
delivering fresh gas to galaxies.

Radial metallicity gradients are a promising probe of inflows 
because they are expected to flatten once previously-ejected gas begins 
to re-accrete.  This is because metals that form and are ejected in a 
galaxy's core are ``spun up" by the halo and re-distributed to large 
radii.  To date, however, radial metallicity gradients in nearby galaxies 
have not been demonstrated to require inflows.  Instead, they have been 
attributed either
to radial gradients in ISM properties such as the strength of outflows,
or to a radially-varying ``evolutionary state" in which annuli at large 
radii are simply less evolved.  These conflicting interpretations point
to the need for further theoretical inquiry into how low-redshift 
observations ought to be interpreted.

At high redshifts, simulations predict that metallicity gradients may have 
been stronger than in the local Universe, possibly because of the weaker 
role of recycled gas and mergers at early times.  Meanwhile, observations seem 
split between lensed galaxies with generally strong gradients and unlensed 
galaxies with weak ones.  Further observational work is evidently required 
in order to clarify the level of agreement with theoretical models.

Moving from radial gradients to ensemble statistics, the interplay between 
inflows and gas processing leads to a rich phenomenology in which 
$M_*$, $Z$, SFR, and gas fraction are tightly coupled.  The resulting 
correlations are observed out to at least $z=2$, and they are being 
disentangled via a wide variety of theoretical studies.  

Analytical treatments have shown that the observed $M_*$-$Z$-SFR 
correlation follows naturally from the assumption that gas processing is 
nearly in equilibrium
with inflows, and that enrichment is nearly in equilibrium with dilution.  
Models that do not explicitly assume equilibrium tend to recover it once
their parameters are tuned to match observations.  Consequently, 
they can be used to \emph{infer} inflow rates (in addition to other physical 
parameters) from the $M_*$-$Z$-SFR relation.  They have further shown that
there is a tight connection between $Z$, SFR, and gas fraction in the 
sense that tuning the model to match one observable often yields agreement 
with another, free of charge.  Detailed measurement of inflows and outflows 
in high-resolution numerical simulations support the importance of 
quasi-equilibrium behavior governed by outflows whose efficiency decreases
with mass.  Meanwhile, cosmological simulations (and at least one 
semi-analytical model) find that a realistic $M_*$-$Z$-SFR-gas fraction 
relation arises naturally within sufficiently realistic frameworks.  
This ensemble statistical behavior has been interpreted as support for
predominantly quiescent, equilibrium-like galaxy growth.  

The progress that has been made in understanding the $M_*$-$Z$-SFR relation,
while promising, does not yet explain everything.  Even after removing
the dependence of $Z$ on specific star formation rate $\mathrm{SFR}/M_*$, 
the residual observed scatter in $Z$ at a given $M_*$ remains 
substantial~\citep{salim14,salim15}.  If the SFR (or gas fraction)
dependence can largely be accounted for by relaxing the assumption of gas 
processing equilibrium, then does the residual scatter indicate departure 
from enrichment equilibrium or intrinsic scatter in $\eta$ or $Z_0$? This
will be a useful question to take up in future theoretical studies.

In summary, it is now widely-recognized that star-forming galaxies do most 
of their growing in a quiescent mode where the global SFR, $Z$, and gas 
fraction constantly adjust on a relatively short timescale to reflect the 
influence of inflows.  Further study of the flow of gas and metals in 
numerical simulations is required in order to improve our understanding 
of equilibrium growth, but its broad role is now beyond dispute.  From the 
beginning, measurements of metallicities have driven the development of 
this paradigm.  There is no doubt that they will continue to do so.

\begin{acknowledgement}
The author thanks R.\ Dav{\'e} and A.\ Fox for offering him the opportunity
to contribute this chapter, and for their patience as he drafted it.  
Additional thanks go to A.\ Klypin for detailed and honest comments 
on an early version.
\end{acknowledgement}

% References
%%%%%%%%%%%%%%%%%%%%%%%% referenc.tex %%%%%%%%%%%%%%%%%%%%%%%%%%%%%%
% sample references
% %
% Use this file as a template for your own input.
%
%%%%%%%%%%%%%%%%%%%%%%%% Springer-Verlag %%%%%%%%%%%%%%%%%%%%%%%%%%
%
% BibTeX users please use
% \bibliographystyle{}
% \bibliography{}
%

\end{document}